\documentclass[%
 reprint,
nofootinbib,
 amsmath,amssymb,
 aps,
]{revtex4-1}
\usepackage{graphicx}
\usepackage{dcolumn}
\usepackage{bm}
\usepackage{color}
\usepackage[T1]{fontenc}
\usepackage{mathrsfs,amsfonts,dsfont}
\usepackage{amstext}
\usepackage{mathtools}
\usepackage{enumitem}
\graphicspath{{graphics/}}
\newcommand{\Tr}[1]{\text{Tr}\Big[#1\Big]}
\newcommand{\Trt}[1]{\text{Tr}[#1]}
\newcommand{\ket}[1]{\left\vert#1\right\rangle}
\newcommand{\bra}[1]{\left\langle#1\right\vert}

\newcommand{\rep}{\text{rep}}
\newcommand{\click}{\text{click}}
\newcommand{\di}{\text{DI}}
\newcommand{\dd}{\text{DD}}
\newcommand{\sift}{\text{sift}}
\newcommand{\raw}{\text{raw}}
\newcommand{\artanh}[1]{\text{artanh}\big(#1\big)}

\begin{document}
\preprint{APS/123-QED}
\title{
Device-independent secret-key-rate analysis for quantum repeaters
}
\author{Timo Holz}\email{holzt@uni-duesseldorf.de}
\author{Hermann Kampermann}
\author{Dagmar Bru\ss}
\affiliation{Theoretical Physics $\MakeUppercase{\romannumeral3}$, Heinrich Heine University Duesseldorf, D-40225 Duesseldorf, Germany}
\date{\today}
\begin{abstract}
The device-independent approach to quantum key distribution (QKD) aims to establish a secret key between two or more parties with untrusted devices, potentially 
under full control of a quantum adversary. The performance of a QKD protocol can be quantified by the secret key rate, which can be lower bounded via the 
violation of an appropriate Bell inequality in a setup with untrusted devices. We study secret key rates in the device-independent scenario 
for different quantum repeater setups and compare them to their device-dependent analogon. The quantum repeater setups 
under consideration are the original protocol by Briegel \emph{et al.} 
and the hybrid quantum repeater protocol by van Loock \emph{et al.}. 
For a given repeater scheme and a given QKD protocol, the secret key rate depends on a variety of parameters, 
such as the gate quality or the detector efficiency. We systematically analyze the impact of these parameters and suggest optimized strategies.
\end{abstract}
\maketitle
\section{\label{sec:1intro}\protect Introduction}
Quantum cryptography --- the science of (secure) private communication based on fundamental properties of quantum particles --- is a very active field of research 
and was founded in the early $1980$s~\cite{foundationQcrypt}. An unconditionally secure encryption technique, the one-time pad~\cite{onetimepad}, relies on a 
preshared key between the 
parties who wish to communicate. Secure communication can thus be achieved by securely distributing this key, which is the ultimate task of quantum key distribution 
(QKD). The famous BB$84$ protocol~\cite{BB84} was the first proposal for achieving secure QKD. 
Since then, a variety of other QKD protocols have been published~\cite{ekert,b92,6state}. However, the security of these 
\emph{device-dependent} (DD) protocols relies on a perfect characterization of the measurement devices and the source, which is impossible in practice. Any realistic 
implementation is imperfect, which makes these QKD protocols vulnerable to an adversary~\cite{secQKDbrassard,secQKDlutkenhaus,Insecurity2006,Insecurity2011}. 
Ideally, one wants to drop any assumption 
about any device involved in the QKD scheme, which is referred to as \emph{device-independent} (DI) QKD~\cite{scaraniDI,DI1}.\\
As photons possess a long coherence time, one can transmit these particles through fibers or free space, thus allowing long-distance QKD. 
Due to photon losses, though, which exponentially scale with the distance one wants to overcome, QKD is limited to distances 
of $L\lesssim150~\text{km}$~\cite{scarani_secfrac,PLOB}. This problem can be circumvented with quantum repeaters~\cite{oqr}.\\
In this work, we aim at comparing achievable secret key rates in the DD and DI scenario for different quantum repeaters without implemented error 
correction. In particular, we provide a systematic analysis on how experimental quantities and errors manifest themselves in the corresponding secret key rates. 
The DD case has been analyzed in~\cite{first}. Here, we shed light on the fundamental differences between both scenarios, especially the requirements needed for a 
reasonably high DI secret key rate.\\
The structure of this paper is as follows. In Sec.~\ref{sec:2framework} we review a generic quantum repeater model~\cite{oqr}, recapitulate the fundamentals 
of QKD, and explain the peculiarities in the device-independent case. Important ingredients, 
such as the secret key rate $R$ and the errors we account for, are described. In Sec.~\ref{sec:3oqr} we apply the given framework to the 
original quantum repeater proposal by Briegel \emph{et al.}~\cite{oqr}. 
Section~\ref{sec:4hqr} focuses on the key analysis for the hybrid quantum repeater~\cite{hqr}.
\section{\label{sec:2framework} General framework}
The main source of errors in quantum communication with photons are 
losses in the optical fiber, which scale exponentially with the length $L_0$, such that the transmittivity $\eta_t$ is given by
\begin{eqnarray}
 \eta_t(L_0)=10^{-\alpha\frac{L_0}{10}} \label{transmittivity}, 
\end{eqnarray}
where $\alpha$ denotes the attenuation coefficient. In this work we use $\alpha=0.17~\text{dB}\slash \text{km}$, which is the attenuation coefficient at 
wavelengths around $1550~\text{nm}$. To overcome the exponential photon loss, quantum repeaters for long-distance quantum information transmission have been suggested.\\
In this section we review a generic model for a quantum repeater, originally introduced by Briegel \emph{et al.}~\cite{oqr}. 
Furthermore, we briefly discuss other sources of errors in QKD and how we model and incorporate them in the quantum repeater scheme. 
See~\cite{first} for a detailed discussion of imperfections. We also review the main ideas of DIQKD, in particular the DI protocol that we use~\cite{scaraniDI}.
\subsection{\label{ssec:2.1genericqrep} Generic quantum repeater model}
\begin{figure}[h!]
\centering
\includegraphics[width=0.375\textwidth]{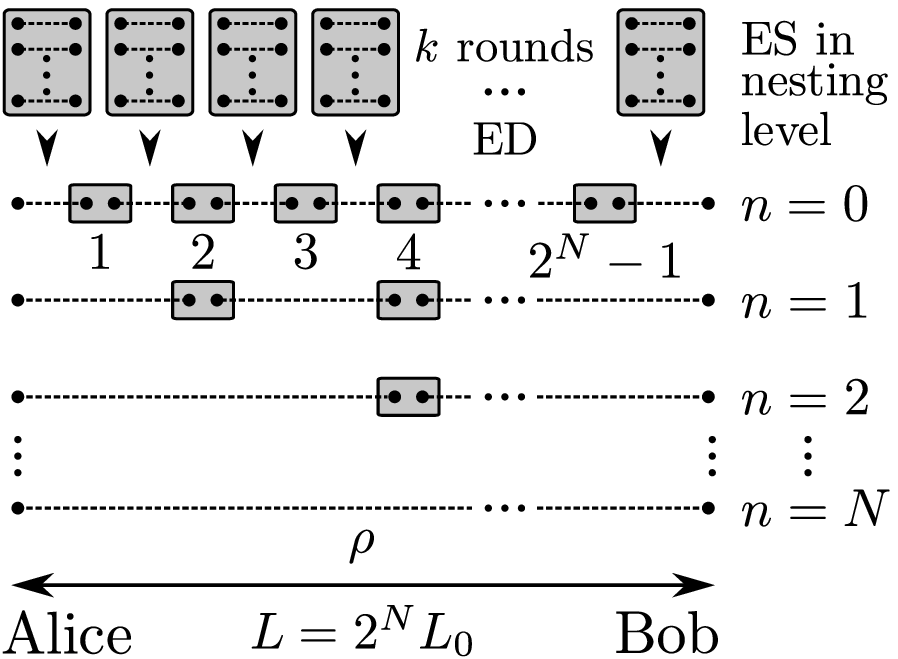}
\caption{A generic quantum repeater setup, proposed by~\cite{oqr}. Let $k$ denote the number of distillation rounds performed prior to the first ES and $N$ the 
maximum nesting level. Alice and Bob are separated by the distance $L=2^NL_0$ and share at the end of the nested protocol the entangled state $\rho$.}
\label{fig:genericqrep}
\end{figure}

The purpose of a quantum repeater is to generate and distribute entangled states over a large distance $L$ that separates two parties, typically called Alice and Bob. 
In order to 
increase the distance over which the states are entangled, one performs entanglement swapping (ES) at intermediate repeater stations equally separated by a 
fundamental length $L_0$. In the nested quantum repeater proposal (see Fig.~\ref{fig:genericqrep} 
for a schematic representation), ES is performed in $N$ consecutive nesting levels, where 
$2^N$ segments of fundamental length $L_0$ amount to the total distance $L=2^NL_0$, which corresponds to $2^N-1$ intermediate repeater stations. For the sake of 
simplicity, we only allow state purification via entanglement distillation (ED) before the first ES is done. The repeater stations are equipped with 
quantum memories and processors to perform the mentioned quantum operations. 
For ED, we employ the Deutsch \emph{et al.}~\cite{deutsch} protocol, which generates after $k$ rounds of distillation a final state of high purity out 
of $2^k$ copies of an initial state $\rho_0$. The ES protocol involves Bell measurements, which can be implemented in various ways in the 
experiment~\cite{ESimplementation, weinfurter}. We review the ED and ES protocol in Appendix~\ref{sec:app0}.\\
As entanglement can be used as a resource for many quantum informational tasks~\cite{superdense,teleportation}, 
it is important to quantify the number of entangled states that can be distributed between Alice and Bob per second by a quantum repeater. This 
quantity is described by the repeater rate $R_\rep$, which clearly depends on errors that occur in the quantum repeater. We briefly discuss which errors are 
taken into account and how we model them. Afterwards we discuss the time restrictions that we focus on and explicitly give the expression 
for the repeater rate.
\subsubsection{Errors of the quantum repeater}
The elements of a quantum repeater and their errors are as follows: 
(i)~Quantum channel~-- Photon losses in the fiber are described via the transmittivity $\eta_t$, Eq.~\eqref{transmittivity}. 
(ii)~Source~-- We assume that the source creates on demand a state $\rho_0$ and distributes it  
to adjacent repeater stations. The quality of these states is described via the fidelity $F_0$, with respect to a certain Bell state, defined in Eqs.~\eqref{Bell12} 
and~\eqref{Bell34}. 
(iii)~Detectors~-- We assume photon number resolving detectors (PNRDs) with efficiency $\eta_d$, 
where dark counts of the detectors are neglected. This is a reasonable approximation for realistic dark counts of the order of $10^{-5}$ or below, see~\cite{first}.
(iv)~Gates~-- ED and ES rely on controlled two-qubit operations, implemented by a gate with quality $p_G$. This imperfect gate introduces noise, thus mixing the ideal pure entangled state. 
We further assume that one-qubit gates work perfectly.\\
The errors in (i)--(iv) give rise to a success probability for ED in round $k$ and for ES in nesting level $n$. We denote those probabilities 
with $P_\text{ED}^{(k)}$ and $P_\text{ES}^{(n)}$, respectively. Finally, let $P_0$ denote the probability that a source successfully links two adjacent repeater 
stations in the $0$th nesting level with an initial entangled state $\rho_0$.
\subsubsection{Repeater rate}
For a given set of parameters and within a model that respects the errors we introduced in the previous section, one can achieve a certain repeater rate $R_\rep$. 
In order to characterize this repeater rate, we need to clarify which time restrictions we account for. 
The only time-consuming operation that we consider is the time needed to distribute an entangled photon 
pair among adjacent repeater stations and acknowledge their successful transmission. This so-called fundamental time $T_0$ 
depends on the speed of light $c=2\times10^8~\text{m}/\text{s}$ in the fiber, the fundamental length $L_0$ separating two repeater stations, and the location 
of the photon source. We consider the case where the source is located at one repeater station, which yields the fundamental time $T_0=2L_0/c$~\cite{first}. 
Furthermore, we investigate repeaters with deterministic and probabilistic ES, i.e., $P_\text{ES}^{(n)}=1$ and $P_\text{ES}^{(n)}<1$, respectively.
\noindent\paragraph{Deterministic ES.} For perfect detectors $\eta_d=1$, the ES can be performed in a deterministic manner. The corresponding repeater rate is given 
by~\cite{deterministic_reprate}
\begin{align}
 R_\rep^\text{det}=\frac{1}{T_0} \frac{1}{Z_n\big(P_{L_0}^{(k)}\big)}, \label{det_reprate}
\end{align}
where the recursive probability $P_{L_0}^{(k)}$ in distillation round $k$ is defined via
\begin{align}
P_{L_0}^{(k)}\coloneqq\frac{P_\text{ED}^{(k)}}{Z_1\big(P_{L_0}^{(k-1)}\big)} \quad\forall\,\, k\ge1 \label{recursive_prob}
\end{align}
and $P_{L_0}^{(0)}\coloneqq P_0$. Here, $Z_n(p)$ denotes the average number of attempts to successfully establish $2^n$ entangled pairs (each generated with 
probability $p$) and it is given by~\cite{deterministic_reprate}
\begin{align}
 Z_n(p)\coloneqq\sum\limits_{j=1}^{2^n}\binom{2^n}{j}\frac{(-1)^{j+1}}{1-(1-p)^j}.
\end{align}
The $2^n$ generated pairs are then deterministically converted via ES in the repeater stations to an entangled pair between Alice and Bob.
\noindent\paragraph{Probabilistic ES.} ES is a probabilistic procedure for imperfect detectors. Given $P_0\ll1$, the repeater rate of a quantum repeater with $k$ 
rounds of ED and ES in $n$ nesting levels can be approximated by
\begin{align}
 R_\rep^\text{prob}=\frac{1}{T_0}\left(\frac{2}{3}\right)^{n+k}P_0\prod\limits_{j=1}^k\frac{P_\text{ED}^{(j)}}{a_\text{ED}^{(j-1)}}
 \prod\limits_{i=1}^n\frac{P_\text{ES}^{(i)}}{a_\text{ES}^{(i-1)}},
 \label{prob_reprate}
\end{align}
which is a generalized and slightly modified version of the repeater rates given in~\cite{gisin,first}.\footnote{In~\cite{gisin}, 
the repeater rate for probabilistic ES is derived without initial ED and without the constants 
$a_{\text{ES}}^{(i)}$. In~\cite{first} initial ED is included and a common constant $a_\text{ED}$ is introduced for every ED round, 
which results in a larger repeater rate. In general, it is not justified to use a common constant $a_\text{ED}$, as they quickly approach unity for an increasing 
number of ED steps. As we show in Appendix~\ref{sec:applast}, one can tackle this problem in a more efficient way and one can similarly introduce constants for 
the ES procedure.} 
Here, $a_\text{ED}^{(j)}$ and $a_\text{ES}^{(i)}$ denote constants that one has to choose depending on success probabilities to create an entangled state in 
the corresponding ED round and nesting level, respectively. They fulfill $0<a_\text{ED}^{(j)},a_\text{ES}^{(i)}\le1$ and are typically close to $1$. 
The repeater rate in Eq.~\eqref{prob_reprate} underestimates the actual repeater rate, 
as already pointed out in~\cite{deterministic_reprate}. Recently, a more sophisticated approach to quantify the 
repeater rate with probabilistic ES appeared in the literature~\cite{LoockNew}.\footnote{Note, however, that for 
more than $n=2$ nesting levels, the repeater rate of~\cite{LoockNew} rapidly becomes only numerically feasible and provides no further insight into our analysis. 
Also, since we want to keep $n$ in principle arbitrary, we settle for the approximated repeater rate in Eq.~\eqref{prob_reprate}.
} To our knowledge, an analytical study of the optimal strategy has not been performed yet.\footnote{
In practice, the optimal strategy for maximizing the repeater rate is to immediately perform ES as soon as entangled pairs are available in two neighboring 
repeater links and then proceed by already distributing new states among these available repeater stations. Monte Carlo simulations suggest that this approach can 
significantly exceed the analytical repeater rates in Eqs.~\eqref{det_reprate} and~\eqref{prob_reprate}, depending on $n$.
}
\subsection{\label{ssec:2.2qkd} Quantum key distribution}
With the repeater rates in Eqs.~\eqref{det_reprate} and~\eqref{prob_reprate}, we now study the possibility to use the entangled states as a resource 
to generate a secret key.
\subsubsection{Device-dependent QKD}
\begin{figure}[h!]
\centering
\includegraphics[width=0.4\textwidth]{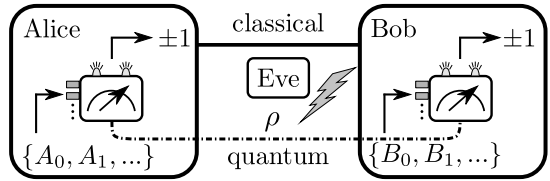}
\caption{A typical QKD setup. Alice and Bob share a classical and a quantum channel. 
A source provides possibly entangled states $\rho$ that can be measured by the 
perfectly characterized measurement devices. A dichotomic classical output is generated in each measurement round.}
\label{fig:ddqkd}
\end{figure}

Suppose that Alice and Bob share a classical, authenticated channel and a possibly entangled state $\rho$, transmitted through a quantum channel. A typical QKD setup is 
shown in Fig.~\ref{fig:ddqkd}. 
In each measurement round, Alice and Bob can choose from a set of measurement settings $\{A_0,A_1,\dots\}$ and $\{B_0,B_1,\dots\}$. 
The setting determines which measurement is performed on their subsystem. 
Throughout this work we consider dichotomic measurement outcomes $a_i,b_j\in\{\pm1\}$.\\
The performance of a QKD protocol is quantified by the secret key rate~\cite{first}
\begin{align}
R\coloneqq R_\raw r_\infty=R_\rep R_\sift P_\click r_\infty, \label{seckeyrate}
\end{align}
which is our figure of merit. The quantities introduced in Eq.~\eqref{seckeyrate} are the 
raw key rate $R_\raw$, the fraction $R_\sift$ of measurements performed in the same basis by Alice and Bob, the probability $P_\click$ for a valid measurement 
result, and the secret fraction $r_\infty$ (see below).\\
After generating an arbitrarily long bit string, the classical postprocessing of the measurement data begins, including sifting, which corresponds to discarding 
measurements where the settings of Alice and Bob did not match. Note that we fix $R_\sift=1$, which can be approximately achieved by choosing the measurement 
settings with biased probabilities~\cite{asymmetricBB84}. 
The sifted or raw key leads to the raw key rate $R_\raw$, which is the number of raw bits Alice and Bob generate per second. 
These bits are only partially secure, which is described by the secret fraction $r_\infty$. The explicit form of 
$r_\infty$ depends on the protocol one employs. A variety of QKD protocols exists in the literature, such as the BB$84$ and the six-state protocol~\cite{BB84,6state}. 
In these QKD protocols one has full knowledge about the Hilbert space dimensions, which is crucial for 
the security of these protocols. For instance, the security of the BB$84$ protocol critically depends on the four dimensions of the Hilbert space associated to 
a qubit pair~\cite{BB84unsecure}. The secret fraction for the BB$84$ protocol is given by~\cite{scarani_secfrac}
\begin{align}
 r_\infty^{\text{BB}84}=\max\big\{0,1-h(Q_z)-h(Q_x)\big\}. \label{secfracDD}
\end{align}
In Eq.~\eqref{secfracDD} the binary entropy is denoted as $h(p)\coloneqq-p\log_2(p)-(1-p)\log_2(1-p)$ and the 
quantum bit error rate (QBER) in measurement direction $i$ is $Q_i$. The QBER is defined as the probability that Alice and Bob generate discordant outcomes, 
given a fixed set of measurement settings, i.e., 
\begin{subequations}
\begin{align}
Q_{z}=P(a\neq b\mid A=Z, B=Z) , \label{Qz}\\
Q_{x}=P(a\neq b\mid A=X, B=X) \label{Qx}
\end{align}
\end{subequations}
for measuring Pauli $Z$ and $X$ operators.
\subsubsection{\label{ssec:2.3diqkd} Device-independent QKD}
\begin{figure}[h!]
\centering
\includegraphics[width=0.4\textwidth]{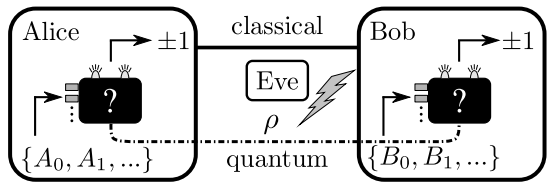}
\caption{The DIQKD setup. 
The measurement devices are treated as black boxes, i.e., the exact internal operations are unknown. Additionally, the dimension of the 
Hilbert space associated to the state $\rho$ is not specified.}
\label{fig:diqkd}
\end{figure}

In practice, it is impossible to have full control over the devices involved in a QKD setup. 
The idea of DIQKD is to extract a secret key without making detailed 
assumptions about the involved devices~\cite{scaraniDI}. 
The security of such DIQKD protocols is based on a loophole-free Bell-inequality violation~\cite{FakingViolation}, for which we have to assume 
that the two parties are causally separated. In the spirit of device independence, the measurement devices are treated as black boxes that perform some 
(unknown) measurement conditioned on a classical input chosen by Alice and/or Bob. The measurement should again yield a dichotomic classical output. 
However, in practice sometimes detectors fail and produce no outcome. 
Measurements where any of the black boxes do not produce an output have to be incorporated into the measurement data.  
Alice and Bob can achieve this by randomly assigning a measurement result $\{\pm1\}$ to such events~\cite{MeasurementResultGuessing}. 
In this sense, every event is a valid DIQKD measurement, yielding $P_\click=1$. Note that these events can be incorporated in our description by substituting the 
final state that Alice and Bob share in the following way:
\begin{align}
 \rho \quad \to \quad \eta_d^2\rho+\frac{1-\eta_d^2}{4}\mathds{1}, \label{subs}
\end{align}
where $\eta_d$ refers to the probability that a no-detection event was replaced by a random outcome. 
Note that $\eta_d$ enters the expression in~\eqref{subs} quadratically, because two detectors of the same efficiency are involved in each 
measurement. Figure~\ref{fig:diqkd} 
shows the DIQKD setup. The DI secret key rate can be calculated via 
\begin{align}
 R^\di=R_\raw r_\infty^\di=R_\rep r_\infty^\di,\label{seckeyrateDI} 
\end{align}
where we used $P_\click=1$ and $R_\sift=1$ (see above). In the DD case, the probability $P_\click$ is a function of the detector efficiency $\eta_d$,
whereas in the DI scenario $\eta_d$ enters the secret fraction $r_\infty^\di$ due to the modification of the quantum state in~\eqref{subs}.\\
Comparing Eqs.~\eqref{seckeyrate} and~\eqref{seckeyrateDI} reveals that both key rates share the common repeater 
rate $R_\rep$, which is consistent with the fact that the purpose of the quantum repeater is simply to provide entangled states 
to the two parties. Alice and Bob can then choose to trust their devices or not. 
Several DIQKD protocols have been proposed in the literature~\cite{scaraniDI,vaziraniDI,pawlowskiDI}. We employ the protocol in~\cite{scaraniDI}.
\subsubsection{DIQKD protocol}
In the DIQKD protocol of~\cite{scaraniDI} Alice randomly (with biased probabilities) chooses between three measurement settings $\{A_0,A_1,A_2\}$. 
The exact internal measurement process is unknown, but the device generates a 
dichotomic classical output $a\in\{\pm1\}$ (no-detection events get an assignment of $\pm1$, uniformly at random). Similarly, Bob 
chooses between two measurement settings $\{B_0,B_1\}$, 
producing a binary output $b\in\{\pm1\}$ in each round. 
A random small subset of their (classical) measurement data generated with the setting $\{A_2,B_1\}$ is 
used to estimate $Q\coloneqq P(a\neq b\mid A_2,B_1)$ and the outcomes of the settings $\{A_{0\slash1},B_{0\slash1}\}$ are used to calculate 
\begin{align}
 S\coloneqq\Tr{\rho\!\!\!\sum\limits_{i,j\in\{0,1\}}^{}(-1)^{i\cdot j}A_i\otimes B_j}. \label{violation}
\end{align}
The main result of~\cite{scaraniDI} is a lower bound for the DI secret fraction of the remaining measurement data of the setting $\{A_2,B_1\}$, given by 
\begin{align}
r_\infty^\di =\max\Big\{0,1-h(Q)-h\Big(\frac{1+\sqrt{S^2/4-1}}{2}\Big)\Big\}, \label{secfracDI}
\end{align}
under the condition that $S>2$ and that the marginal probabilities of Alice and Bob are symmetric, i.e., 
$\Trt{\rho A_i\otimes\mathds{1}}=0=\Trt{\rho\mathds{1}\otimes B_j}$ for all $i,j$. 
This lower bound was proven for collective attacks and one-way classical postprocessing in~\cite{scaraniDI}. 
See also~\cite{SimpleAndTight} for more general quantum adversaries and general communication between the parties. 
In the following section we adopt the specific implementation given in~\cite{scaraniDI}, where $Q$ and $S$ are the QBER and 
the Clauser-Horne-Shimony-Holt (CHSH) parameter~\cite{chsh}, respectively.
\subsubsection{\label{sssec:comparison}Comparing DDQKD and DIQKD protocols}
To point out the distinct features separating both scenarios and how they impact the secret key rates, we have to make the DD and the DI protocol effectively 
comparable. The specific implementation given in~\cite{scaraniDI} for the DI protocol uses
\begin{subequations}
\label{implementationDI}
\begin{eqnarray}
A_{0,1}&=\frac{ X\pm Z}{\sqrt{2}}, \quad &A_2= Z , \\
B_0&= X,\quad &B_1= Z. \label{DIdirectionsBob}
\end{eqnarray}
\end{subequations}
for the measurement operators. To compare this to the BB$84$ protocol, where Alice uses $\{A_x=X,A_z=Z\}$ and Bob $\{B_0,B_1\}$ as in 
Eq.~\eqref{DIdirectionsBob}, we also consider the asymmetric implementation of the DI protocol, such that 
$\{A_2=Z,B_1=Z\}$ is measured with probability $\to1$ and with a negligible, but equal fraction with which the other 
measurement operators are used. In the DI and DD case they use these measurement settings to estimate the CHSH value, Eq.~\eqref{violation}, and the QBER $Q_x$, 
respectively. Then, in the asymptotic limit, these protocols are equivalent in the sense that almost always the $Z$-measurement is used. 
Alice and Bob only rely on different assumptions regarding the trust in their measurement devices.
\subsubsection{\label{sssec:state_model} Entangled state, QBER and CHSH parameter}
The explicit form of the state that is distributed to Alice and Bob by the quantum repeater is of fundamental importance for achievable secret key rates. 
Maximal correlation, and thus maximal security is provided if the state $\rho$ is pure and in one of the four Bell states:
\begin{subequations}
\begin{align}
\ket{\phi_{1,2}}&\coloneqq\frac{1}{\sqrt{2}}\big(\ket{00}\pm\ket{11}\big), \label{Bell12} \\ 
\ket{\phi_{3,4}}&\coloneqq\frac{1}{\sqrt{2}}\big(\ket{01}\pm\ket{10}\big). \label{Bell34}
\end{align}
\end{subequations}
For the specific implementation in Eqs.~\eqref{implementationDI}, the ideal state is the pure state $\ket{\phi_1}$ 
for which the CHSH parameter reaches its maximum value $2\sqrt{2}$~\cite{tsirelson} and the QBERs vanish. 
Then, the DD and DI secret fraction are both equal to $1$, which maximizes the corresponding secret key rates. In practice, the source cannot 
provide perfectly pure states due to noise and other imperfections. Under the assumption that the initially distributed states $\rho_0$ 
are genuine two-qubit states, they can be transformed into a generic Bell-diagonal state
\begin{align}
\rho_0=\sum_{i=1}^4c_{i,0}^{(0)}\ket{\phi_i}\bra{\phi_i} \label{generic_bell}
\end{align}
by using local operations~\cite{generic2}.\footnote{Note that depolarizing reduces only nonlocal correlations.} 
The Bell coefficients $c_{i,0}^{(0)}$ are non-negative and fulfill normalization $\sum_ic_{i,0}^{(0)}=1$. We assume throughout this work that the sources generate 
the generic Bell-diagonal state given in Eq.~\eqref{generic_bell}. The ED and ES protocols we use produce Bell-diagonal states, provided 
the input states have been of the form~\eqref{generic_bell}. The quantum repeater thus distributes the final state,
\begin{align}
 \rho=\sum_{i=1}^4c_{i,n}^{(k)}\ket{\phi_i}\bra{\phi_i}, \label{final_bell}
\end{align}
to Alice and Bob, where $c_{i,n}^{(k)}$ denotes the Bell coefficients after ED in $k$ rounds and ES in $n$ nesting levels. 
The coefficients $c_{i,n}^{(k)}$ fulfill normalization, and they depend on $c_{i,0}^{(0)}$ and on the explicit form of the protocol. 
See~\cite{deutsch,first} or Appendix~\ref{sec:app0} for details of the protocols. The transformation rules for the coefficients $c_{i,n}^{(k)}$ under ED and ES 
are summarized in Appendixes~\ref{sec:app1} and~\ref{sec:app2} for the two quantum repeater setups. For Bell-diagonal states, as in Eq.~\eqref{final_bell}, the QBERs 
$Q_{x,n}^{(k)}$ and $Q_{z,n}^{(k)}$ are given by
\begin{subequations}
 \label{QxQzDD}
\begin{align}
Q_{x,n}^{(k)}&=c_{2,n}^{(k)}+c_{4,n}^{(k)} ,\\ 
Q_{z,n}^{(k)}&=c_{3,n}^{(k)}+c_{4,n}^{(k)} .
\end{align}
\end{subequations}
To calculate the quantities needed for the DI secret fraction, one needs to substitute the state $\rho$, Eq.~\eqref{final_bell}, with its noisy version~\eqref{subs}. 
This results in 
\begin{subequations}
\label{QzSDI}
\begin{align}
Q_{z,n}^{(k)}&= \eta_d^2\big(c_{3,n}^{(k)}+c_{4,n}^{(k)}\big)+\frac{1-\eta_d^2}{2},\\
S_n^{(k)}&= 2\sqrt{2}\eta_d^2\big(c_{1,n}^{(k)}-c_{4,n}^{(k)}\big),
\end{align}
\end{subequations}
where $S_n^{(k)}$ denotes the violation of the CHSH inequality with the final state.
\section{\label{sec:3oqr}The Original Quantum Repeater}
Now we want to compare achievable secret key rates for the original quantum repeater (OQR)~\cite{oqr} in the DD and DI scenario.
In Sec.~\ref{ssec:missing_oqr} we give the missing expressions needed to calculate the repeater rate $R_\text{rep}$. 
This is followed by a systematic secret-key-rate analysis, 
where we compare the DD and DI QKD performance numerically (Sec.~\ref{ssec:performance_oqr}) and analytically (Sec.~\ref{ssec:analytical}). 
Since any two-qubit mixture can be transformed into depolarized Bell states with local operations~\cite{divincenzo}, 
we assume that the sources initially distribute such states with Bell coefficients $c_{1,0}^{(0)}=F_0$ and 
$c_{i\ge2,0}^{(0)}=(1-F_0)\slash3$, where $F_0$ denotes the fidelity with respect to the Bell state $\ket{\phi_1}$. 
\subsection{\label{ssec:missing_oqr}Parameters and error model}
In order to calculate the repeater rate $R_\text{rep}$, we need to specify the probabilities 
$P_0$, $P_\click$, $P_\text{ES}^{(n)}$, and $P_\text{ED}^{(k)}$ and how the gate quality $p_G$ enters the expression. 
The probability that the source successfully connects two adjacent repeater stations with an entangled photon pair is given by the 
transmittivity $P_0=\eta_t(L_0)$, Eq.~\eqref{transmittivity}, and the probability for a valid QKD measurement is $P_\click=\eta_d^2$. 
The ED and ES protocol employ controlled two-qubit gates that may introduce noise due to imperfections. We adopt the depolarizing model 
of~\cite{oqr} for noisy gates,
\begin{align}
 \mathcal{O}(\chi)=p_G\mathcal{O}^\text{ideal}(\chi)+\frac{1-p_G}{4}\mathds{1}, \label{pGoqr}
\end{align}
where $\chi$ denotes an arbitrary two-qubit state on which the gate $\mathcal{O}$ acts. 
The ED and ES include twofold detections with PNRDs of efficiency $\eta_d$. 
For perfect detectors $\eta_d=1$, the repeater rate is given by Eq.~\eqref{det_reprate}. In case of nonperfect detectors, however, 
the detection events lead to a factor $\eta_d^{2(k+n)}$ for the success probabilities $P_\text{ED}^{(k)}$ and $P_\text{ES}^{(n)}$. 
Starting from Eq.~\eqref{prob_reprate}, we thus get
\begin{align}
 R_\rep^\text{prob}=\frac{1}{T_0}\left(\frac{2}{3}\right)^{k+n}\eta_d^{2(k+n)}\eta_t(L_0) \prod\limits_{i=1}^{n}\frac{1}{a_\text{ES}^{(i-1)}}
 \prod\limits_{j=1}^k\frac{P_\text{ED}^{\prime\,(j)}}{a_\text{ED}^{(j-1)}} \label{reprateOQR}
\end{align}
for the repeater rate with probabilistic ES, where $P_\text{ED}^{\prime\,(j)}$ now denotes the success probability for ED in round $j$ without the detector efficiency 
$\eta_d$, which can be calculated via the coefficients $c_{i,0}^{(j)}$ only (see Appendix~\ref{sec:app1}, Eq.~\eqref{EDsuccess_oqr}).
\subsection{\label{ssec:performance_oqr}Performance: DD vs DI secret key rate}
With the framework provided in the previous sections, we now want to systematically analyze achievable secret key rates in the DD and DI scenario. 
We split the analysis into two parts, one with perfect detectors $\eta_d=1$ and one with imperfect detectors $\eta_d<1$, as this quantity determines which 
repeater rate has to be used for the calculation. Currently feasible PNRDs reach detector efficiencies of $\eta_d\approx0.95$ at wavelengths 
around $1550~\text{nm}$~\cite{PNRDs}. 
\subsubsection{Perfect detectors}
\begin{figure}[h!]
\centering
\includegraphics[width=0.475\textwidth]{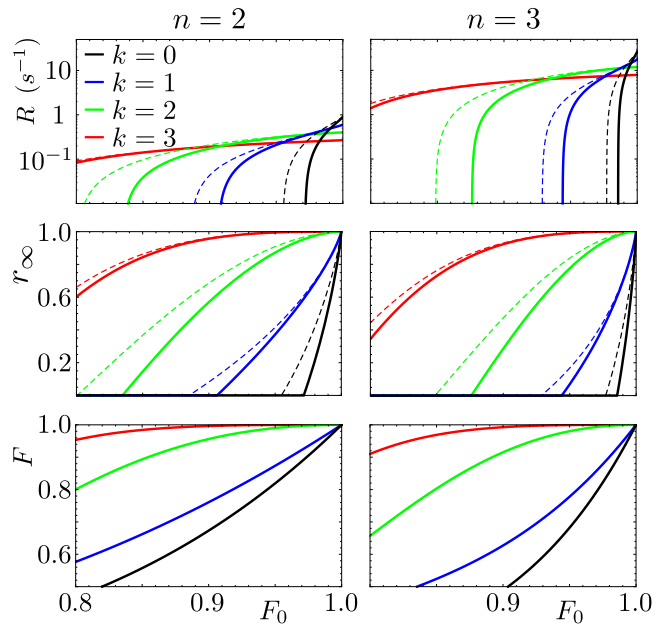}
\caption{Secret key rate $R$, secret fraction $r_\infty$, and final fidelity $F=c_{1,n}^{(k)}$ with respect to $\ket{\phi_1}$ in the DD 
(dashed lines) and DI (solid lines) scenario versus the initial fidelity $F_0$ for the gate quality $p_G=1$, perfect detector efficiency 
$\eta_d=1$, and the total distance $L=600~\text{km}$. 
Different numbers of initial ED rounds are shown, where $k=0$ corresponds to the rightmost curve, and $k=3$ to the leftmost one. 
The left (right) column represents $n=2$ ($n=3$) nesting levels, which corresponds to a fundamental 
length $L_0=150~\text{km}$ ($L_0=75~\text{km}$). Note that for $\eta_d=1$, the curves for the final fidelity $F$ for the DD and DI scenario coincide.}
\label{fig:p1n23}
\end{figure}

\begin{figure}[h!]
\centering
\includegraphics[width=0.475\textwidth]{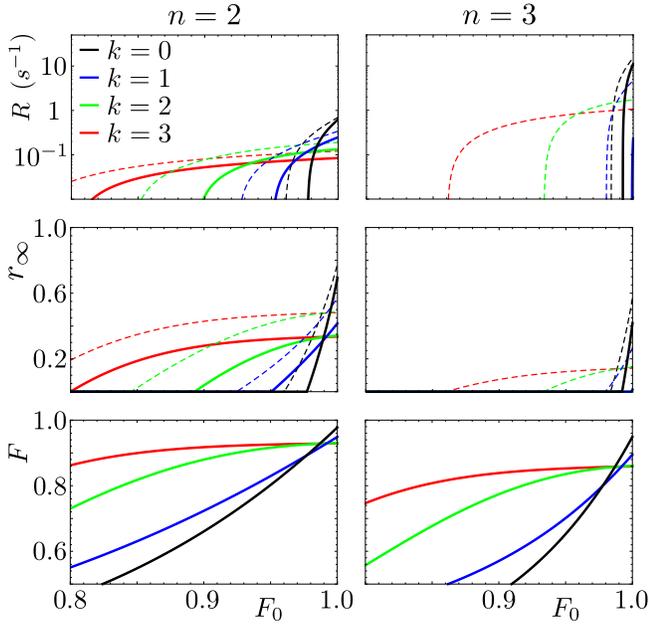}
\caption{Secret key rate $R$, secret fraction $r_\infty$, and final fidelity $F=c_{1,n}^{(k)}$ with respect to $\ket{\phi_1}$ in the DD 
(dashed lines) and DI (solid lines) scenario versus the initial fidelity $F_0$ for $p_G=0.99$, perfect detector efficiency $\eta_d=1$, and the total 
distance $L=600~\text{km}$. Different numbers of initial ED rounds are shown, where $k=0$ corresponds to the rightmost curve, and $k=3$ to the leftmost one. 
The left (right) column represents $n=2$ ($n=3$) nesting levels. Note that for $\eta_d=1$, the curves for the final fidelity $F$ for the DD and DI scenario 
coincide.}
\label{fig:p0995n23}
\end{figure}

For this part we use the deterministic repeater rate in Eq.~\eqref{det_reprate}. 
Note that for $\eta_d=1$, the differences in the secret key rates solely originate from the DD and DI secret fraction. 
We begin the performance analysis with perfect gate qualities $p_G=1$ to understand how ED 
and ES influence the secret key rates. Suppose Alice and Bob are separated by the total distance $L=600~\text{km}$. 
At the end of the repeater protocol, they receive a Bell-diagonal state with coefficients $c_{i,n}^{(k)}$. Figure~\ref{fig:p1n23} 
shows the secret key rates $R$ and $R^\di$ (upper subfigures), the corresponding secret fractions $r_\infty$ and $r_\infty^\di$ (middle subfigures), 
and the fidelity $F(\ket{\phi_1},\rho)\coloneqq\bra{\phi_1}\rho\ket{\phi_1}$ of the final state $\rho$ and the pure Bell state $\ket{\phi_1}$ 
(lower subfigures) as a function of the initial fidelity $F_0$ for various numbers of initial ED rounds $k$ and nesting levels $n$. 
The secret key rates are calculated via Eqs.~\eqref{seckeyrate} and~\eqref{seckeyrateDI}. The secret fractions, Eqs.~\eqref{secfracDD} and~\eqref{secfracDI}, 
are calculated via the QBERs and the CHSH parameter given in Eqs.~\eqref{QxQzDD} and~\eqref{QzSDI}.\\
The first feature that one notices is the fact that $R\ge R^\di$ holds, which is what we expect, since in the DD case, Alice and Bob can rely on more assumptions, 
which directly leads to a higher secret fraction. This should hold in any fair DD to DI comparison. The secret key rates are only identical in the 
ideal case where $\eta_d=1$, $p_G=1$, and $F_0=1$. Only under 
these perfect conditions do Alice and Bob share the pure and maximally entangled state $\ket{\phi_1}\bra{\phi_1}$, which yields a secret fraction of $1$. Comparing 
the case of $n=2$ nesting levels with $n=3$, one observes that both secret key rates significantly increase with $n$. For perfect gates, it is advantageous 
to reduce the fundamental length $L_0$ to decrease photon losses. This holds although more intermediate 
repeater stations involve more noisy states connected by ES, which reduces the secret fractions $r_\infty$ and $r_\infty^\di$ as shown in Fig.~\ref{fig:p1n23}. 
For a larger number of ED rounds $k$, both QKD protocols become more resistant to noise in the initial state $\rho_0$ but they suffer from an overall smaller 
secret key rate, as several copies of states are consumed. 
From the lower subfigures, we observe that ED and ES are two counteracting processes, when it comes to the final fidelity $F$ with respect to $\ket{\phi_1}$. 
This is consistent with the shown secret fractions, since a lower fidelity $F$ results in an increase of the QBERs and in a decrease of the CHSH parameter 
(see Eqs.~\eqref{QxQzDD} and~\eqref{QzSDI}).\\
We now consider imperfect gates. Figure~\ref{fig:p0995n23} 
shows the same quantities as in Fig.~\ref{fig:p1n23} but for $p_G=0.99$. The lower gate quality has a 
strong impact on the DI secret fraction $r_\infty^\di$ and thus also on the DI secret key rate, especially for more nesting levels $n$. 
The mixing of the final state due to noisy gates has a significantly larger influence on the CHSH parameter as it has on the QBER $Q_x$. 
If the source distributes states with a high initial fidelity $F_0$, it is not beneficial for the final fidelity $F$ to perform any ED. 
(See crossing points of solid lines in Fig.~\ref{fig:p0995n23}.)
\subsubsection{Imperfect detectors}
\begin{figure}[h!]
\centering
\includegraphics[width=0.475\textwidth]{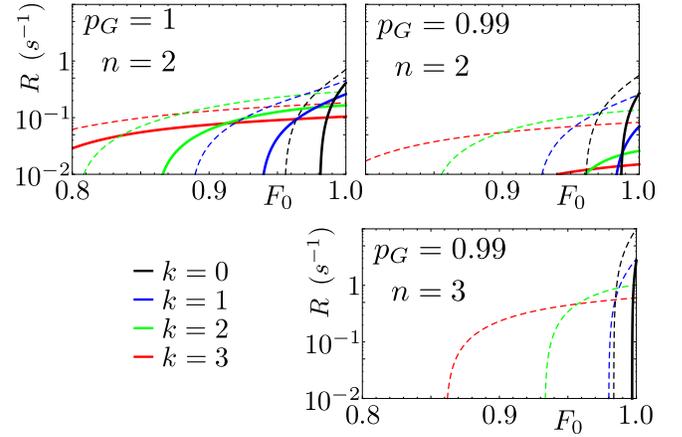}
\caption{DD (dashed lines) and DI (solid lines) secret key rate versus the fidelity $F_0$ with imperfect detectors of efficiency 
$\eta_d=0.975$ and the total distance $L=600~\text{km}$. 
We use different gate qualities $p_G$ and different 
number of nesting levels $n$. The rightmost curve corresponds to $k=0$ and the leftmost curve to $k=3$ initial ED rounds.}
\label{fig:Roqr}
\end{figure}

\begin{figure}[h!]
\centering
\includegraphics[width=0.475\textwidth]{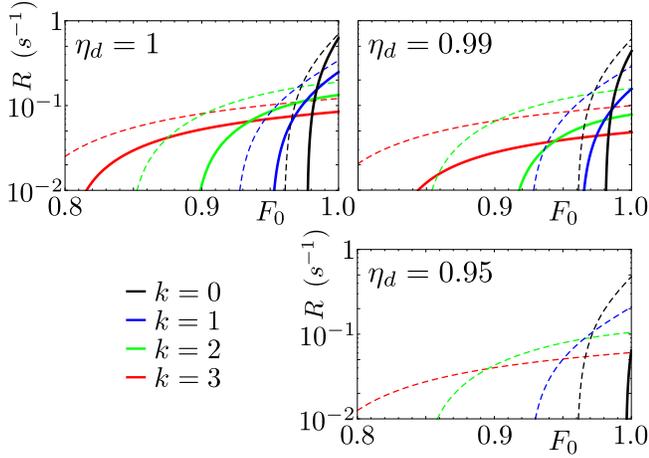}
\caption{DD (dashed lines) and DI (solid lines) secret key rate versus the fidelity $F_0$ for various different detector 
efficiencies $\eta_d$. The gate quality, the number of nesting levels, and the total length are set to 
$p_G=0.99$, $n=2$, and $L=600~\text{km}$, respectively. The rightmost curve corresponds to $k=0$ and the leftmost curve to $k=3$ initial ED rounds.
} 
\label{fig:Roqr_etaimpact}
\end{figure}

For an imperfect detector efficiency $\eta_d<1$, the repeater rate is calculated via Eq.~\eqref{prob_reprate}. 
The DD secret key rate additionally suffers from the global scaling factor $P_\click=\eta_d^2$ (see Eq.~\eqref{seckeyrate}). 
In the DI scenario, however, the lack of perfect detectors is equivalent to performing QKD with states having increased noise, see substitution~\eqref{subs}. 
These differences aside, the DD and DI secret key rates can be calculated as before. Fig.~\ref{fig:Roqr} 
compares the secret key rates as a function of the fidelity $F_0$ for various numbers of ED rounds $k$, different numbers of nesting levels $n$, and 
different gate qualities $p_G$ for $\eta_d=0.975$ and $L=600~\text{km}$. 
By comparing the upper two subfigures, we again observe that the gate quality has a much stronger impact on the DI secret key rate. 
Reducing $p_G=1$ to $p_G=0.99$ results in significantly smaller DI secret key rates, while the DD secret key rates 
are more or less of the same order. The difference between the DD and DI secret key rate becomes higher by increasing the number of initial ED rounds, 
which indicates that the number of imperfect quantum operations is a critical quantity for DIQKD. This is also confirmed by the lower subfigure, where we increased 
the number of nesting levels from $n=2$ to $n=3$. 
One gets only a nonvanishing DI secret key rate for $k=0$, whereas the DD secret key rates 
gain about $1$ order of magnitude. Recall that performing ES in more nesting levels decreases the fundamental length $L_0$, thus 
reducing the probability of photon losses in the fiber. This explains the higher DD secret key rates for $n=3$. However, in the DI case, the errors introduced by 
imperfections outweigh the benefits that one gains from a reduced fundamental length $L_0$. 
Hence, in the DI case one has to accept a larger amount of photon losses in the fiber of larger fundamental length $L_0$ in comparison to the DD case. 
In addition, one has to ensure that the source distributes entangled states of high initial fidelity $F_0$. 
This decreases the number of ED and ES steps and thus reduces the errors introduced by imperfect devices. We conclude that in general, the 
strategy for optimizing the DI secret key rate is different from the DD case.\\
In Fig.~\ref{fig:Roqr_etaimpact} 
we vary the detector efficiency $\eta_d$ and keep the gate quality $p_G$ fixed. It compares DI (solid lines) and DD (dashed lines) 
secret key rates for various values of $\eta_d$ and confirms the intuition that a reduction of the detector efficiency has a larger impact on the DI secret key rate.  
We observe a similar pattern as in Fig.~\ref{fig:Roqr}. With a decreasing detector efficiency both secret key rates drop, but the DI secret key rate is far more 
affected by the imperfections of the detector than its DD analogon.
\subsection{Analytical results - Performance \label{ssec:analytical}}
As the secret fractions are calculated via the coefficients $c_{i,n}^{(k)}$ of the final Bell-diagonal state, 
it is desirable to analytically characterize the behavior of the coefficients $c_{i,n}^{(k)}$ under ED and ES operations with imperfect devices. Formulating general 
analytical results is cumbersome due to the recursive nature of the transformation rules for the 
Bell coefficients under ED and ES, see Eqs.~\eqref{EDtrafo_oqr} and~\eqref{EStrafo_oqr}. In an idealized scenario, where the source distributes pure states, 
however, we can find closed transformation rules for the coefficients $c_{i,n}^{(k)}$, depending on the number of nesting levels $n$ and the gate quality $p_G$. 
We thus consider the case $c_{1,0}^{(0)}=F_0=1$ and $c_{i\ge2,0}^{(0)}=0$, and since ED is obsolete for maximally entangled states we set $k=0$. 
One can show via Eqs.~\eqref{EStrafo_oqr} that the coefficients transform according to 
\begin{align}
 c_{1,n}^{(0)}=\frac{1+3p_G^{\bar{n}}}{4} \!\quad\text{and}\quad \! c_{i\ge2,n}^{(0)}=\frac{1-p_G^{\bar{n}}}{4} \quad \forall\,\,n\in\mathbb{N}, \label{simplified_trafo}
\end{align}
where $\bar{n}\coloneqq2^n-1$ denotes the number of intermediate repeater stations. With Eq.~\eqref{simplified_trafo} one can express the QBERs and the 
CHSH parameter in terms of $\bar{n}$ and $p_G$. For the DD QBERs, Eqs.~\eqref{QxQzDD}, one immediately finds
\begin{align}
Q_{x,n}^{(0)} = Q_{z,n}^{(0)} = \frac{1-p_G^{\bar{n}}}{2} \label{QxQz_simplified}
\end{align}
and for the DI quantities via Eqs.~\eqref{QzSDI} similarly,
\begin{subequations}
\begin{align}
 Q_{z,n}^{(0)} &= \frac{1-\eta_d^2p_G^{\bar{n}}}{2}, \\
 S_n^{(0)} &= 2\sqrt{2}\eta_d^2p_G^{\bar{n}}.\label{QzS_simplified}
\end{align}
\end{subequations}
Recall that the DI secret fraction is only nonvanishing if the CHSH inequality is violated. Thus, we obtain the condition
\begin{align}
 S_n^{(0)}>2 \quad \Leftrightarrow \quad \eta_d^2p_G^{\bar{n}}>\frac{1}{\sqrt{2}}, \label{condition}
\end{align}
which the parameters $p_G$, $\eta_d$, and $\bar{n}$ have to fulfill. The DD and DI secret fractions then become
\begin{subequations}
\label{boiled_down}
\begin{align}
r_\infty^\dd&=\eta_d^2\Big[1-2h\Big(\frac{1-p_G^{\bar{n}}}{2}\Big)\Big], \label{rDD_boiled_down}\\
r_\infty^\di&=1-h\Big(\frac{1-\eta_d^2p_G^{\bar{n}}}{2}\Big)-h\Big(\frac{1}{2}+\frac{1}{2}\sqrt{2\eta_d^4p_G^{2\bar{n}}-1}\Big),\label{rDI_boiled_down}
\end{align}
\end{subequations}
where for $r_\infty^\dd$, we included the factor $\eta_d^2$ compared to Eq.~\eqref{secfracDD}. 
Now, we can investigate the impact of the experimental quantities 
$\eta_d$, $p_G$, and $n$ onto the secret fractions in terms of partial derivatives, which are given in Eqs.~\eqref{rDDderivatives} and~\eqref{rDIderivatives} 
in Appendix~\ref{ssec:app1.2}. We quantify the influence of the parameter onto the secret fractions via these partial derivatives and 
thus ask the question which of the two secret fractions, DD or DI, alters its value faster when the corresponding parameter is changed.
\subsubsection{Impact of the detector efficiency $\eta_d$}
Using the fact that 
$\partial_{\eta_d}r_\infty^\di$ is a monotonic function and respecting the condition given in Eq.~\eqref{condition}, one can show that the inequality 
$\partial_{\eta_d}r_\infty^\di>\partial_{\eta_d}r_\infty^\dd$ holds, see Eq.~\eqref{ordering_etaimpact} in Appendix~\ref{ssec:app1.2} for details. 
Hence, the DI secret fraction reacts more sensitively to changes in the detector efficiency than the effective DD secret fraction does.
\subsubsection{Impact of the gate quality $p_G$}
\begin{figure}[h!]
\centering
\includegraphics[width=0.475\textwidth]{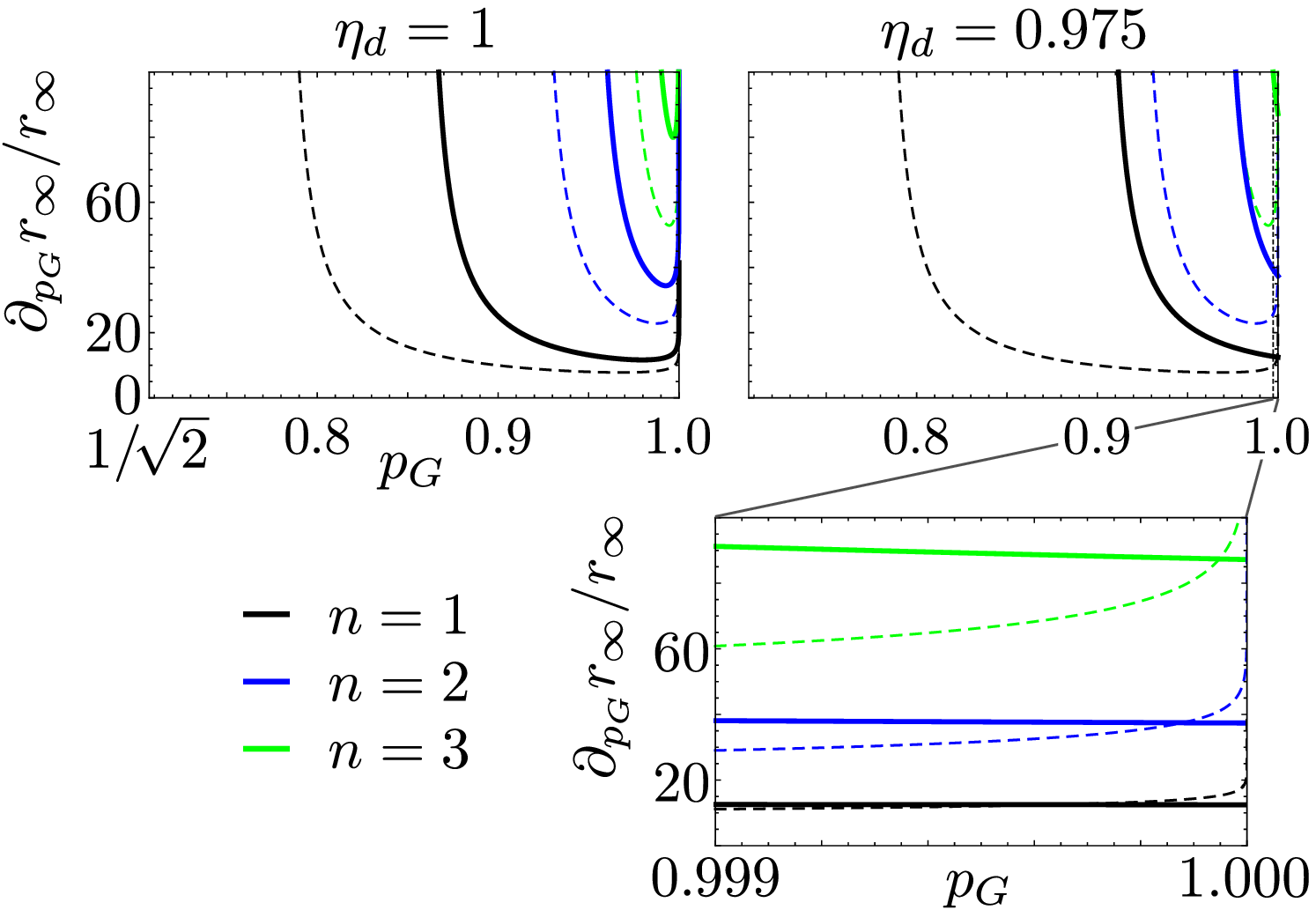}
\caption{Relative change $\partial_{p_G}r_\infty\slash r_\infty$ versus the gate quality $p_G$ in the DD (dashed lines) and DI 
(solid lines) scenario for detector efficiencies $\eta_d=1$ and $\eta_d=0.975$ (see Eqs.~\eqref{rDDderivatives_b} and~\eqref{rDIderivatives_b}). 
Different numbers of nesting level $n$ are shown, where $n=1$ corresponds to the leftmost curves and $n=3$ to the 
rightmost ones.
}
\label{fig:ratio_p}
\end{figure}

For the derivatives of the secret fractions with respect to the gate quality $p_G$ and the nesting levels $n$, 
the ordering of the corresponding expressions in Eqs.~\eqref{rDDderivatives} and~\eqref{rDIderivatives} in Appendix~\ref{ssec:app1.2} is not as obvious as for the 
detector efficiency $\eta_d$. Thus, for the sake of simplicity, we settle for a numerical comparison. Figure~\ref{fig:ratio_p} 
shows the relative change of the derivatives $\partial_{p_G}r_\infty$ in the DD (Eq.~\eqref{rDDderivatives_b}) and DI (Eq.~\eqref{rDIderivatives_b}) case 
with respect to the corresponding secret fraction $r_\infty$ for $\eta_d=1$ and $\eta_d=0.975$ as a function of the gate quality. 
We observe that the relative change of the DI secret fraction is larger than its DD analogon. 
For $\eta_d<1$ and almost perfect gates $1-p_G\ll1$, though, the opposite is true (see inset in Fig.~\ref{fig:ratio_p}). 
This follows from the fact that $\partial_{p_G}r_\infty^\di$ no longer diverges for $p_G\to1$ and $\eta_d<1$, in contrary to $\partial_{p_G}r_\infty^\dd$; see 
Eqs.~\eqref{rDDderivatives_b} and~\eqref{rDIderivatives_b}.\\
However, an important difference is that the relative change in the DI case also depends on the 
detector efficiency $\eta_d$, in contrast to the DD case. Figure~\ref{fig:ratio_p} also verifies the intuition that the impact of the gate quality $p_G$ rises with 
an increasing number of nesting levels, i.e., with an increasing number of imperfect quantum operations.
\subsubsection{Impact of the nesting levels $n$} 
\begin{figure}[h!]
\centering
\includegraphics[width=0.475\textwidth]{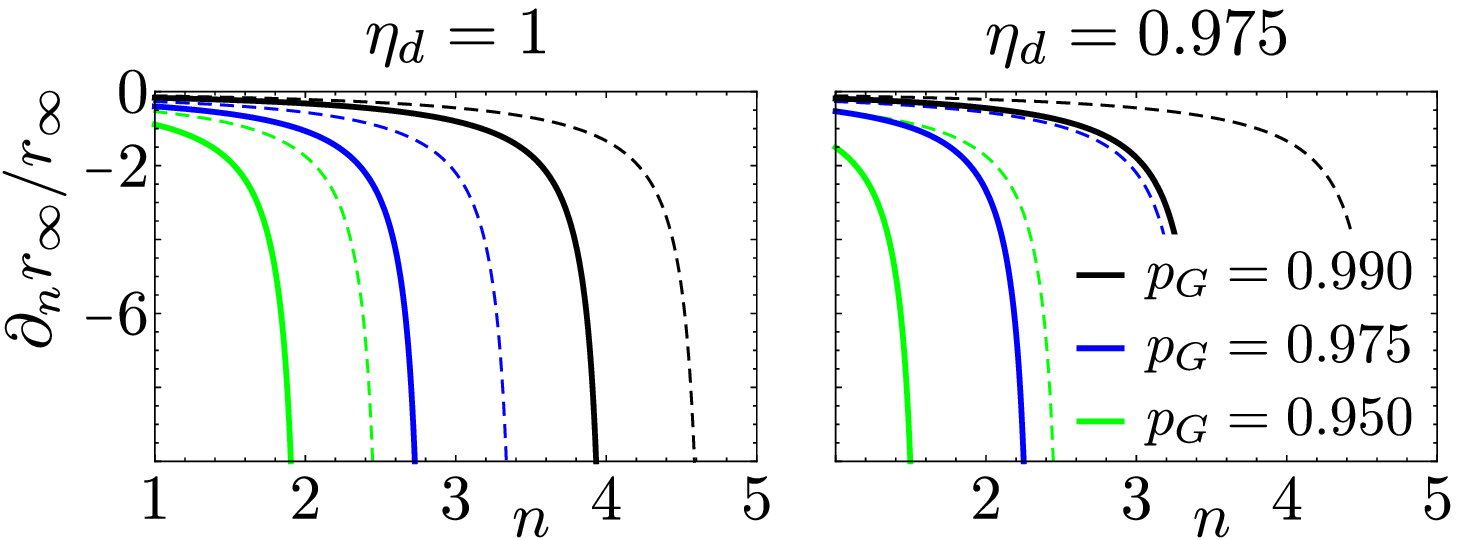}
\caption{Relative change $\partial_{n}r_\infty\slash r_\infty$ versus the number $n$ in the DD (dashed lines) and DI 
(solid lines) scenario for detector efficiencies $\eta_d=1$ and $\eta_d=0.975$ (see Eqs.~\eqref{rDDderivatives_c} and~\eqref{rDIderivatives_c}). The 
rightmost curves correspond to the gate quality $p_G=0.99$ and the leftmost ones to $p_G=0.95$.
}
\label{fig:ratio_n}
\end{figure}

To quantify the influence of $n$, let us extrapolate the integer $n$ to a continuous variable. In Fig.~\ref{fig:ratio_n} 
we numerically compare the relative change of $\partial_nr_\infty$, Eqs.~\eqref{rDDderivatives_c} 
and~\eqref{rDIderivatives_c}, with respect to corresponding secret fractions $r_\infty$. 
It confirms that the relative change $\partial_nr_\infty\slash r_\infty$ in the DI case is larger than its DD analogon, as expected. 
Note that $\partial_nr_\infty\slash r_\infty$ is negative and that the 
DD ratio is again independent of the detector efficiency $\eta_d$. One can also observe that the impact of 
$n$ dramatically increases with a decreasing gate quality $p_G$, which is consistent with previous results.\\
To close this section we conjecture that our analytical results approximately hold for sufficiently pure 
initial states, since $\epsilon$ small contributions to other Bell states $\ket{\phi_{i\neq1}}$ in the initially distributed states do not 
significantly alter the state at the end of the ES protocol.
\section{\label{sec:4hqr}The Hybrid Quantum Repeater}
Let us now consider the hybrid quantum repeater (HQR) introduced by van Loock \emph{et al.}~\cite{hqr} and Ladd 
\emph{et al.}~\cite{hqr2}. It still employs the nested scheme for ES as shown in Fig.~\ref{fig:genericqrep}, but the repeater stations and the physical system 
representing the qubits are of fundamental difference compared to the OQR. As in \cite{first}, we also restrict our investigation to HQRs where 
unambiguous state discrimination (USD) measurements are involved for state generation~\cite{hqr_add1,hqr_add2}. 
In Part~\ref{ssec:missing_hqr} of this section, we introduce the concepts of HQRs, and in Part~\ref{ssec:performance_oqr} the comparison of the 
DD-DI performance follows.
\subsection{\label{ssec:missing_hqr}Setup, error model and repeater rate}
In Sec.~\ref{sssec:hqr1} we review the model for intermediate repeater stations and briefly capture the main ideas behind the entanglement creation in 
this setup. Afterwards, we present in Sec.~\ref{sssec:hqr2} the error model for noisy two-qubit gates and explain how to calculate the 
repeater rate. See~\cite{first} for more details.
\subsubsection{\label{sssec:hqr1}Repeater station - Model}
\begin{figure}[h]
\centering
\includegraphics[width=0.475\textwidth]{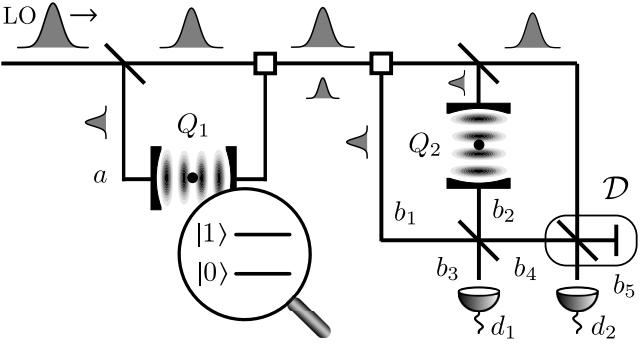}
\caption{Illustration of repeater stations in the HQR setup and the USD scheme following~\cite{hqr_add2}. 
A coherent state $\ket{\alpha}$, the local oscillator (LO), is generated and sent through linear optical 
elements, such as beam splitters and optical switches. Different optical modes are denoted with $a$ and $b_i$, for $i\in\{1,\dots,5\}$. The LO passes a beam 
splitter and a part of it interacts with qubit $Q_1$, which is prepared in an equally weighted superposition of its two possible states $\ket{0}$ and 
$\ket{1}$. The resulting optical state is sent together with the LO to the next repeater station, where again a part of 
the LO interacts with qubit $Q_2$, also prepared in an equally weighted superposition of $\ket{0}$ and $\ket{1}$. 
A $50:50$ beam splitter is applied to modes $b_1$ and $b_2$ and a displacement operation $\mathcal{D}$ to the pulse in mode 
$b_4$. Depending on the measurement results of detectors $d_1$ and $d_2$, an entangled state between qubits $Q_1$ and $Q_2$ is generated.}
\label{fig:setup_hqr}
\end{figure}

The HQR combines discrete and continuous degrees of freedom. Entanglement is for instance generated between two trapped ions inside a cavity, which represent the qubits. 
The entangling interaction, however, is induced via coherent optical states. The interaction between the qubits and the light can thus be described within the 
Jaynes-Cummings framework~\cite{JC}. A schematic model for intermediate repeater stations is shown in Fig.~\ref{fig:setup_hqr}. 
By performing a USD measurement on the optical modes, after they interacted with the qubits, the entangled state
\begin{align}
 \rho_0=F_0\ket{\phi_1}\bra{\phi_1}+(1-F_0)\ket{\phi_2}\bra{\phi_2} \label{rho0hqr}
\end{align}
can be conditionally prepared. For the HQR, the probability $P_0$ to connect two adjacent repeater stations with an entangled state is given by~\cite{first}
\begin{align}
 P_0=1-(2F_0-1)^{\frac{\eta_t\eta_d}{1+\eta_t(1-2\eta_d)}}. \label{p0hqr} 
\end{align}
Note that the probability $P_0$ vanishes for pure states $\rho_0=\ket{\phi_1}\bra{\phi_1}$ with $F_0=1$, in 
which case it is not possible to generate a secret key. For more details regarding the implementation and state preparation see~\cite{first,hqr_add2}.
\subsubsection{\label{sssec:hqr2}Error model and repeater rate}
ES and ED rely on controlled-$Z$ operations. The model for a noisy two-qubit gate needs to be adjusted for the HQR implementation. 
According to~\cite{errorhqr}, the noisy two-qubit gate $\mathcal{O}$ acting upon the 
two-qubit state $\chi\equiv\chi_{\text{ab}}$, which describes the main errors due to dissipation, is modeled by
\begin{align}
\mathcal{O}(\chi)=\mathcal{O}^\text{ideal}\Big(&p_c^2(p_G)\chi+(1-p_c(p_G))^2 Z_{\text{a}} Z_{\text{b}}\rho Z_{\text{a}}Z_{\text{b}} \nonumber \\ 
&\!\!\!\!\!+p_c(p_G)(1-p_c(p_G))(Z_{\text{a}}\chi Z_{\text{a}}+Z_{\text{b}}\chi Z_{\text{b}})\Big).
\label{pGhqr}
\end{align}
Here,
\begin{align}
 p_c(p_G)\coloneqq
 \frac{1+\exp\left(-\frac{\pi(1-p_G^2)}{2\sqrt{p_G}(1+p_G)}\right)}{2} \label{quantities_errorhqr}
\end{align}
represents the probability for each qubit to not suffer a $Z$ error. The quantity $p_G$ in Eq.~\eqref{quantities_errorhqr} is the local transmission 
parameter that describes the effect of photon losses onto the gate and can thus be seen as an effective gate quality. 
Following~\cite{first}, we calculate the repeater rate according to Eq.~\eqref{det_reprate} with deterministic ES, i.e., $P_\text{ES}=1$. 
We use perfect qubit measurements for the ES and also ED operations, since the imperfections can in principle be eliminated from the protocol at the cost of 
additional photon losses in the quantum channel, which effectively reduces the gate quality~\cite{hqr_add1}. 
Note, however, that we account for detector imperfections at the initial entanglement distribution (as $\eta_d$ enters the probability $P_0$ in 
Eq.~\eqref{p0hqr}) and 
detector imperfections at the final qubit measurements in the laboratories of Alice and Bob. The latter one implies again a factor $P_\click=\eta_d^2$ for the 
DD secret key rate, while in the DI scenario the substitution~\eqref{subs} has to be performed. 
The DD and DI secret fractions are calculated according to Eqs.~\eqref{secfracDD} and~\eqref{secfracDI}, and since the final state is again Bell diagonal, 
the QBERs and the CHSH parameter are given by Eqs.~\eqref{QxQzDD} and~\eqref{QzSDI}.
\subsection{\label{ssec:performance_hqr}Performance: DD vs DI secret key rate}
\begin{figure}[h]
\centering
\includegraphics[width=0.475\textwidth]{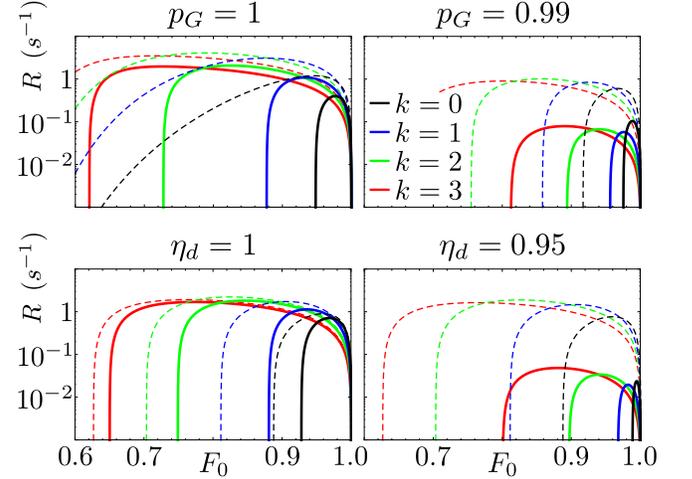}
\caption{DD (dashed lines) and DI (solid lines) secret key rate for the HQR versus the fidelity $F_0$. 
The total distance is $L=300~\text{km}$, with $n=2$ nesting levels. Different numbers of initial ED rounds $k$ are shown, where the most narrow curves 
correspond to $k=0$ and the most wide ones to $k=3$. 
The upper two subfigures show the impact of the effective gate quality, as it is reduced from 
$p_G=1$ to $p_G=0.99$ with a fixed detector efficiency of $\eta_d=0.975$. The lower subfigures similarly display the influence of the detector efficiency, 
where we reduce it from $\eta_d=1$ to $\eta_d=0.95$ with the fixed parameter $p_G=0.995$.} 
\label{fig:Rhqr0}
\end{figure}

\begin{figure}[h]
\centering
\includegraphics[width=0.475\textwidth]{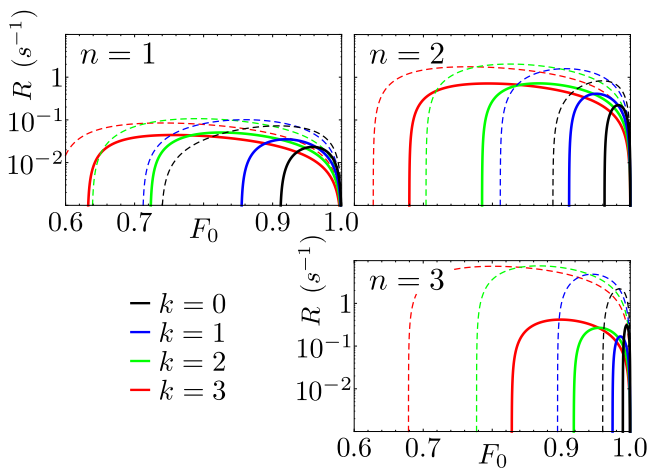}
\caption{DD (dashed lines) and DI (solid lines) secret key rate versus the fidelity $F_0$. 
The total distance, the gate quality, and the detector efficiency are $L=300~\text{km}$, $p_G=0.995$, and $\eta_d=0.975$, respectively. 
As in Fig.~\ref{fig:Rhqr0}, the most narrow curves correspond to $k=0$ and the most wide ones to $k=3$ ED rounds. The figure shows the impact of different nesting 
levels $n$, varied from $n=1$ to $n=3$.} 
\label{fig:Rhqr}
\end{figure}

We now want to investigate the influence of the effective gate quality $p_G$, the detector efficiency $\eta_d$, and the 
number of ED and ES operations on the secret key rates. Figure~\ref{fig:Rhqr0}
shows the DD and DI secret key rates versus the fidelity $F_0$ for several numbers of initial ED rounds $k$. The total distance is $L=300~\text{km}$ 
with a fixed number of nesting levels $n=2$. 
We can observe from the upper two subfigures that gate imperfections have a large impact on the DD secret key rate, as already pointed out in~\cite{first}. 
In the DI case, this becomes even more dramatic. The lower two subfigures show that detector errors do not significantly reduce the DD secret key rate. 
The DI secret key rate, however, is heavily compromised by these imperfections, as they lead to a mixed state due to the random assignment of measurement results.\\
We conclude the key rate analysis with Fig.~\ref{fig:Rhqr}, 
where the secret key rates are shown as a function of the initial fidelity $F_0$ for several numbers of nesting levels $n$ at a fixed total distance of 
$L=300~\text{km}$. We consider gate and detector errors by $p_G=0.995$ and $\eta_d=0.975$, respectively. 
As we can see, it is beneficial for the DD secret key rate to increase the number of nesting levels beyond $n=2$ 
to reduce photon losses in the fiber. By doing so, the DD secret key rates gain approximately $1$ order of magnitude. 
In the DI case, however, the errors introduced by the larger number of imperfect quantum operations 
outweigh again the benefits that one gains from a reduced fundamental length $L_0$. For a given fidelity $F_0$ the optimal number $k$ of ED rounds is in general 
different from the DD scenario as well. 
\section{\label{sec:5conclusion}Conclusion and Outlook}
In this work, we provided a detailed systematic analysis on achievable secret key rates of two quantum repeater setups in the device-independent (DI) scenario and 
compared it to the device-dependent (DD) case. We studied the original quantum repeater (OQR)~\cite{oqr} and the hybrid quantum repeater 
(HQR)~\cite{hqr}. The analysis includes a numerical investigation on how experimental quantities, such as the gate quality $p_G$, the detector efficiency $\eta_d$, 
the initial fidelity $F_0$, and the number of nesting levels $n$ and initial entanglement distillation rounds $k$, influence the secret key rate. 
We observed for both setups that the DI security comes at the expense 
of being particularly sensitive towards malfunctions in the devices. Imperfections of the gates, the detectors, and the sources compromise the achievable DI secret key 
rate more than the DD one. Hence, for any realistic implementation, there is a gap between these secret key rates that increases with an increasing number of 
imperfect quantum operations. For the OQR with an idealized photon source, we additionally verified analytically that the parameters $p_G$, $\eta_d$, and $n$ have a 
stronger impact on the DI secret key rate, as they have in the DD scenario.\\
The proneness of DIQKD to imperfections naturally implies different optimization strategies for the DI and DD secret key rate. 
In the DD scenario the influence of the gate errors is not as severe as it is in the DI case, thus allowing a shorter fundamental distance $L_0$ and thus 
reducing photon losses in the fiber, i.e., in the DI case there are not as many intermediate repeater stations feasible as in the DD one. 
This immediately yields a stronger limitation for the total distance $L$ that one can overcome in the DI setup. Similarly, the purity of the initially distributed states 
can be improved via more entanglement distillation rounds in the DD protocol, which makes it more robust to imperfections of the source.\\
It remains for future investigations to compare different DD and DI protocols, besides the BB$84$ and the modified Ekert protocol \cite{BB84,scaraniDI}. 
Other ideas are to extend this analysis to different quantum repeater models, such as the 
DLCZ quantum repeater~\cite{dlcz}. One could also include more error sources of the quantum repeater, e.g. errors introduced by quantum memories, 
and investigate their impact on the secret key rates. For the latter one, we conjecture from the provided secret-key-rate analysis that further imperfections have a 
qualitatively similar impact on the DI secret key rate as the ones discussed in this work.
\begin{acknowledgments}
The authors acknowledge support from the Federal Ministry of Education and Research (BMBF, Project Q.com-Q) and thank Peter van Loock for discussion.
\end{acknowledgments}
\appendix
\section{\label{sec:applast}Repeater rate: Probabilistic ES} 
Here, we provide more details for the repeater rate with probabilistic ES in Eq.~\eqref{prob_reprate}. In~\cite{gisin}, the repeater rate 
\begin{equation}
 R_\rep^\text{prob}=\frac{1}{T_0}\left(\frac{2}{3}\right)^{n}P_0^\prime\prod\limits_{i=1}^nP_\text{ES}^{(i)} \label{rrep_probES_app}
\end{equation}
without initial ED is derived for $P_0^\prime\ll1$, where $P_0^\prime$ denotes 
the success probability to connect two adjacent repeater stations in nesting level $n=0$ with an entangled pair (see also Fig.~\ref{fig:genericqrep}). 
We review the derivation of Eq.~\eqref{rrep_probES_app} and explain how to improve this rate. Afterwards we include initial ED, inspired by~\cite{first}. 
\subsubsection{Repeater rate without ED}
Following~\cite{gisin}, the number of attempts $n_0$ to successfully create an elementary link is governed by the probability distribution
\begin{equation}
 p(n_0)=(1-P_0^\prime)^{n_0-1}P_0^\prime , \label{probdistribution_n0}
\end{equation}
which yields the expectation value 
\begin{equation}
 \left\langle n_0\right\rangle=\sum\limits_{n\in\mathbb{N}_0}^{}n_0p(n_0)=\frac{1}{P_0^\prime}. \label{expvalue_n0}
\end{equation}
In order to perform ES, one needs entangled states in two neighboring segments of the repeater line. The corresponding combined 
probability distribution is given by
\begin{align}
 \tilde{p}(n_0)=p(n_0)^2+2p(n_0)\sum\limits_{k=1}^{n_0-1}p(k) \label{combined_p_n0},
\end{align}
which results in the average number of attempts 
\begin{align}
 \left\langle\tilde{n}_0\right\rangle=\sum\limits_{n_0\in\mathbb{N}_0}n_0\tilde{p}(n_0)=\frac{3-2P_0^\prime}{(2-P_0^\prime)P_0^\prime}. \label{expvalue_tilde_n0}
\end{align}
The first ES step can now be performed, which succeeds with probability $P_\text{ES}^{(1)}$, thus increasing the average number of attempts to create an 
entangled link in nesting level $n=1$ according to
\begin{equation}
\left\langle n_1\right\rangle=\left\langle\tilde{n}_0\right\rangle\sum\limits_{k\in\mathbb{N}_0}(k+1)(1-P_\text{ES}^{(1)})^kP_\text{ES}^{(1)}=
\frac{\left\langle\tilde{n}_0\right\rangle}{P_\text{ES}^{(1)}}. \label{expvalue_n1}
\end{equation} 
From now on, our approach deviates from the one in~\cite{gisin}, where $\left\langle\tilde{n}_0\right\rangle$ in 
Eq.~\eqref{expvalue_tilde_n0} is set to $3\slash2P_0^\prime$, which is a good approximation for $P_0^\prime\ll1$. Here, we 
rewrite Eq.~\eqref{expvalue_tilde_n0} as
\begin{align}
 \left\langle\tilde{n}_0\right\rangle=\frac{3-2P_0^\prime}{(2-P_0^\prime)P_0^\prime}=\frac{1}{P_0^\prime}\frac{3}{2}a_\text{ES}^{(0)}, \label{expvalue_tilde_n0_reexpress}
\end{align}
where we defined
\begin{align}
 a_\text{ES}^{(0)}\coloneqq\frac{1-2P_0^\prime\slash3}{1-P_0^\prime\slash2}. \label{aESdef}
\end{align}
In complete analogy to Eq.~\eqref{expvalue_n0}, the probability $P_1$ to create an entangled link in nesting level $n=1$ is given by the inverse of 
Eq.~\eqref{expvalue_n1}, and we can define an according probability distribution $p(n_1)$ via $P_1$. 
This is in general not true, as the success probability of establishing a link in a higher nesting level $n=i$ in the $n_i$th attempt depends 
on success probabilities of the previous nesting levels~\cite{gisin} and the corresponding probability distribution $p(n_i)$ is not analog to the form 
given in Eq.~\eqref{probdistribution_n0}. However, this modification allows us to obtain the recursion
\begin{align}
\left\langle n_i\right\rangle&= \frac{1}{P_i}=\frac{\left\langle\tilde{n}_{i-1}\right\rangle}{P_\text{ES}^{(i)}} \quad \forall \,\, i\in\mathbb{N} , \label{recursion1} \\
\left\langle\tilde{n}_i\right\rangle&=\frac{3-2P_i}{(2-P_i)P_i}=\frac{1}{P_i}\frac{3}{2}a_\text{ES}^{(i)} \quad \forall \,\, i\in\mathbb{N},\label{recursion2}
\end{align}
if we iterate this argument to arbitrary nesting levels. The constants $a_\text{ES}^{(i)}$ are defined as in Eq.~\eqref{aESdef} with the corresponding probability $P_i$. 
The beginning of the recursion is given in Eqs.~\eqref{expvalue_n0} and~\eqref{expvalue_tilde_n0_reexpress}. Note that this approach also only yields a good approximation 
for $P_0^\prime\ll1$, but this strategy leads to repeater rates which are closer to achievable ones that are calculated with Monte Carlo simulations.\\
With the relations~\eqref{recursion1} and~\eqref{recursion2} we can express the average number of attempts to establish a single entangled link at the maximum nesting 
level $n=N$ as
\begin{align}
\left\langle n_N\right\rangle&=\frac{\left\langle\tilde{n}_{N-1}\right\rangle}{P_\text{ES}^{(N)}}=
\frac{3}{2}\frac{a_\text{ES}^{(N-1)}}{P_\text{ES}^{(N)}}\frac{1}{P_{N-1}}= \dots \nonumber \\ 
&= \left(\frac{3}{2}\right)^N\frac{1}{P_0^\prime}\prod\limits_{i=1}^{N}\frac{a_\text{ES}^{(i-1)}}{P_\text{ES}^{(i)}}. \label{expvalue_nN}
\end{align}
Each attempt lasts the fundamental time $T_0$, thus yielding the repeater rate
\begin{align}
 R_\rep^\text{prob}=\frac{1}{T_0}\left(\frac{2}{3}\right)^NP_0^\prime\prod\limits_{i=1}^{N}\frac{P_\text{ES}^{(i)}}{a_\text{ES}^{(i-1)}}. \label{rrep_upperbound}
\end{align}
\subsubsection{Repeater rate with ED}
In the spirit of~\cite{first}, we now include initial ED, which is performed at each segment at nesting level $n=0$ and which thus only affects the success probability 
$P_0^\prime$. Thus, $P_0^\prime$ is given by the recursively defined probabilities 
$P_0^\prime=P_{L_0}^{(k)}$ for successful ED in $k$ rounds in Eq.~\eqref{recursive_prob}. By plugging the recursive probabilities into each other, one arrives at
\begin{align}
P_{L_0}^{(k)}&=\frac{2}{3}\frac{P_\text{ED}^{(k)}}{a_\text{ED}^{(k-1)}}P_{L_0}^{(k-1)}=\dots=
\left(\frac{2}{3}\right)^kP_0\prod\limits_{j=1}^{k}\frac{P_\text{ED}^{(j)}}{a_\text{ED}^{(j-1)}}, \label{P0tilde}
\end{align}
where we defined the constants $a_\text{ED}^{(j)}$ for ED as in Eq.~\eqref{aESdef}. Replacing $P_0^\prime$ in Eq.~\eqref{rrep_upperbound} with the right-hand side 
of Eq.~\eqref{P0tilde} yields the repeater rate in Eq.~\eqref{prob_reprate}.
\section{\label{sec:app0}ED and ES protocol}
For completeness, we review the ED and ES protocols~\cite{first,deutsch}, which determine together with the 
noisy two-qubit gate models in Eqs.~\eqref{pGoqr} and~\eqref{pGhqr} the transformation of the coefficients $c_{i,n}^{(k)}$ (see Appendixes~\ref{sec:app1} 
and~\ref{sec:app2}). Let $C_{\text{NOT}}^{s\to t}$ denote a controlled-$X$ operation, where 
$s$ and $t$ indicate the source and the target qubit, respectively.
\subsubsection{Entanglement distillation}
Suppose Alice and Bob share the two states $\rho_{a_i,b_i}$ for $i\in\{1,2\}$. The following steps are performed. 
(i) Alice/Bob rotates her/his particles by $+/-\frac{\pi}{2}$ around the $X$ axis in the computational basis $\{\ket{0},\ket{1}\}$. 
(ii) Alice/Bob applies $C_\text{NOT}^{a_1\to a_2}$/$C_\text{NOT}^{b_1\to b_2}$. 
(iii) The state $\rho_{a_2,b_2}$ is measured in the computational basis. Then, if their measurement results coincide, the state 
$\rho_{a_1,b_1}$ has been purified. Otherwise the state is discarded.
\subsubsection{Entanglement swapping}
Suppose the two entangled states $\rho_{a,b}$ and $\rho_{c,d}$ are distributed among two adjacent repeater stations. 
The following algorithm performs ES between these two states. 
(i) A $C_\text{NOT}^{b\to c}$-gate is applied. 
(ii) Qubits $b$ and $c$ are measured in the basis $\{\ket{\pm}\coloneqq(\ket{0}\pm\ket{1})\slash\sqrt{2}\}$ and $\{\ket{0},\ket{1}\}$, respectively. 
(iii) Depending on the measurement outcomes, a single-qubit rotation on qubit $d$ is performed and one obtains the entangled state $\rho_{a,d}$. 
\section{\label{sec:app1}Additional material: OQR}
\subsection{\label{ssec:app1.1}Transformation under ED and ES}
With the discussed error models and the ED and ES protocols, we recall the transformation rules of the coefficients $c_{i,n}^{(k)}$. 
For the OQR, gate errors are modeled according to Eq.~\eqref{pGoqr}. See~\cite{deutsch,first} for details.
\noindent\paragraph{Entanglement distillation.} 
Two copies of the Bell-diagonal state $\rho^{(k-1)}=\sum_{i=1}^4 c_i^{(k-1)}\ket{\phi_i}\bra{\phi_i}$ represent the input states for the ED protocol. 
Provided the ED protocol is successful, one is left with one Bell-diagonal state with the coefficients
\begin{subequations}
\label{EDtrafo_oqr}
\begin{align}
c_1^{(k)}&=\frac{1}{8P_\text{ED}^{\prime\,(k)}}\big[1+p_G^2\big(8c_1^{(k-1)\,2}+8c_4^{(k-1)\,2}-1\big)\big],	\\
c_2^{(k)}&=\frac{1}{8P_\text{ED}^{\prime\,(k)}}\big[1-p_G^2\big(1-16c_1^{(k-1)}c_4^{(k-1)}\big)\big]	,	\\
c_3^{(k)}&=\frac{1}{8P_\text{ED}^{\prime\,(k)}}\big[1+p_G^2\big(8c_2^{(k-1)\,2}+8c_3^{(k-1)\,2}-1\big)\big],	\\
c_4^{(k)}&=\frac{1}{8P_\text{ED}^{\prime\,(k)}}\big[1-p_G^2\big(1-16c_2^{(k-1)}c_3^{(k-1)}\big)\big],	
\end{align}
\end{subequations}
where the success probability of ED round $k$ is
\begin{align}
 P_\text{ED}^{\prime\,(k)}=\frac{1}{2}\big(1+p_G^2\big(2c_{1}^{(k-1)}+2c_{4}^{(k-1)}-1\big)^2\big). \label{EDsuccess_oqr}
\end{align}
\noindent\paragraph{Entanglement swapping.} 
Two qubit pairs, each in the Bell-diagonal state $\rho_{n-1}=\sum_{i=1}^4 c_{i,n-1}\ket{\phi_i}\bra{\phi_i}$, are the input states to the ES 
protocol that includes a probabilistic Bell measurement on two qubits, one of each pair. The two qubits not involved in the Bell measurement are again in a 
Bell-diagonal state with coefficients $c_{i,n}$. The transformation rules are
\begin{subequations}
\label{EStrafo_oqr}
\begin{align}
c_{1,n}&=\frac{1-p_G}{4}+p_G\sum\limits_{i=1}^{4}c_{i,n-1}^2, \\ 
c_{2,n}&=\frac{1-p_G}{4}+2p_G\big(c_{1,n-1}c_{2,n-1}+c_{3,n-1}c_{4,n-1}\big), \\
c_{3,n}&=\frac{1-p_G}{4}+2p_G\big(c_{1,n-1}c_{3,n-1}+c_{2,n-1}c_{4,n-1}\big), \\
c_{4,n}&=\frac{1-p_G}{4}+2p_G\big(c_{1,n-1}c_{4,n-1}+c_{2,n-1}c_{3,n-1}\big), 
\end{align}
\end{subequations}
and the success probability for ES is given by $P_\text{ES}^{(n)}=\eta_d^2$, neglecting dark counts of the detector.
\subsection{\label{ssec:app1.2}Analytical calculations}
\subsubsection{Partial derivatives of secret fractions}
The partial derivatives of $r_\infty^\dd$, Eq.~\eqref{rDD_boiled_down}, with respect to $\eta_d$, $p_G$, and $n$ are given by:
\begin{subequations}
\label{rDDderivatives}
\begin{align}
\partial_{\eta_d}r_\infty^\dd&= 2\eta_d\Big[1-2h\Big(\frac{1-p_G^{\bar{n}}}{2}\Big)\Big], \label{rDDderivatives_a}\\
\partial_{p_G}r_\infty^\dd&= 2\frac{\bar{n}\eta_d^2p_G^{\bar{n}-1}}{\ln(2)}\artanh{p_G^{\bar{n}}}, \label{rDDderivatives_b}\\
\partial_nr_\infty^\dd&=2(\bar{n}+1)\eta_d^2p_G^{\bar{n}}\ln(p_G)\artanh{p_G^{\bar{n}}}, \label{rDDderivatives_c}
\end{align}
\end{subequations}
where we introduced the area hyperbolic tangent 
\begin{align}
\text{artanh}(x)\coloneqq \frac{1}{2}\ln\left(\frac{1+x}{1-x}\right) \quad \forall\,\, x \in (-1,1), \label{areatanh}
\end{align}
which is the inverse tangent hyperbolic function. The partial derivatives of $r_\infty^\di$, Eq.~\eqref{rDI_boiled_down}, 
with respect to $\eta_d$, $p_G$, and $n$ are
\begin{subequations}
\label{rDIderivatives}
\begin{align}
\partial_{\eta_d}r_\infty^\di&= \frac{2\eta_dp_G^{\bar{n}}}{\ln(2)}q(\eta_d,p_G,\bar{n}) , \label{rDIderivatives_a}\\
\partial_{p_G}r_\infty^\di&=\frac{\bar{n}\eta_d^2p_G^{\bar{n}-1}}{\ln(2)} q(\eta_d,p_G,\bar{n}),\label{rDIderivatives_b}\\
\partial_nr_\infty^\di&=(\bar{n}+1)\eta_d^2p_G^{\bar{n}} \ln(p_G) q(\eta_d,p_G,\bar{n}), \label{rDIderivatives_c}
\end{align}
\end{subequations} 
where the function $q(\eta_d,p_G,\bar{n})$ is defined as:
\begin{align}
 q(\eta_d,p_G,\bar{n})\coloneqq&\frac{2\eta_d^2p_G^{\bar{n}}}{\sqrt{2\eta_d^4p_G^{2\bar{n}}-1}}\text{artanh}\Big(\sqrt{2\eta_d^4p_G^{2\bar{n}}-1}\Big)\nonumber\\
 &+\artanh{\eta_d^2p_G^{\bar{n}}}
  \label{qfct}.
\end{align} 
\subsubsection{Comparison: Impact of detector efficiency}
For the partial derivatives of $r_\infty^\dd$ and $r_\infty^\di$ with respect to the detector efficiency, Eqs.~\eqref{rDDderivatives_a} and~\eqref{rDIderivatives_a}, 
one can derive an ordering relation to show that $\eta_d$ has a larger impact in the DI scenario. 
Note that $\partial_{\eta_d}r_\infty^\di$ is positive for all parameters $\eta_d$, $p_G$, and $\bar{n}$ that fulfill the condition~\eqref{condition} and that 
$\eta_d\partial_{\eta_d}r_\infty^\di$ is a strictly monotonically increasing function of $\eta_d^2p_G^{\bar{n}}$. Hence, the following ordering holds:
\begin{align}
 \partial_{\eta_d}r_\infty^\di&\ge
 \eta_d\partial_{\eta_d}r_\infty^\di\ge\lim\limits_{\eta_d^2p_G^{\bar{n}}\to\sqrt{2}^{-1}}^{}\left(\eta_d\partial_{\eta_d}r_\infty^\di\right) \nonumber\\
 &=\frac{\sqrt{2}}{\ln(2)}\big(\artanh{1\slash\sqrt{2}}+\sqrt{2}\big)>2, \label{ordering_etaimpact}
\end{align}
where we used $\text{artanh}(1/\sqrt{2})>0$ and $0\le\eta_d,\ln(2)\le1$. Finally, note that in the DD case, $\eta_d$ enters the effective secret fraction 
$\eta_d^2r_\infty^{\text{BB}84}$ as a factor with $r_\infty^{\text{BB}84}$ given in 
Eq.~\eqref{secfracDD}. This partially derived with respect to $\eta_d$ yields $2\eta_dr_\infty^{\text{BB}84}$, which is upper 
bounded by $2$. This proves the inequality $\partial_{\eta_d}r_\infty^\di>\partial_{\eta_d}r_\infty^\dd$ as claimed in Sec.~\ref{ssec:analytical}.
\section{\label{sec:app2}Additional material: HQR} 
\subsection{Transformation under ED and ES}
Here, we give the transformation relations for the Bell coefficients under ED and ES for the HQR, 
where gate errors enter the calculation via Eq.~\eqref{pGhqr}. See~\cite{first}.
\noindent\paragraph{Entanglement distillation.} We calculate the coefficients after ED round $k$ with 
respect to the coefficients after ED round $k-1$, which we do not label here explicitly for a better overview. 
Also, we suppress the dependency on $p_G$ of $p_c(p_G)$ and introduce the abbreviation $\bar{p}\coloneqq2p_c(p_c-1)$:
\begin{widetext}
\begin{subequations}
 \label{EDtrafo_hqr}
\begin{align}
c_1^{(k)}&=\frac{1}{P_\text{ED}^{(k)}}\big[
\bar{p}^2\big(c_1-c_4\big)\big(c_1-c_4+c_2-c_3\big)+
\bar{p} \big(c_1^{2}+c_4^{2}+(c_1-c_4)^2 -c_1c_3-c_2c_4\big)+c_1^{2}+c_4^{2}\big], \\
c_2^{(k)}&=\frac{1}{P_\text{ED}^{(k)}}\big[\bar{p}^2\big(c_1c_3+(c_2-c_3-c_4)c_4\big)-\bar{p} \big(c_3+c_4\big)c_4+
2(\bar{p}+1)^2c_1c_4-\bar{p}(\bar{p}+1) c_1\big(c_1+c_2\big)\big]	,	\\
c_3^{(k)}&=\frac{1}{P_\text{ED}^{(k)}}\big[\bar{p}^2\big(c_1c_2+c_3c_4\big)+(\bar{p}+1)^2\big(c_2^2+c_3^2\big)-\bar{p}(\bar{p}+1)\big(c_2(c_3+c_4)+(c_1+c_2)c_3\big)
\big], \\
c_4^{(k)}&=\frac{1}{P_\text{ED}^{(k)}}\big[\bar{p}^2\big(c_2c_4+(c_1-c_3-c_4)c_3\big)-\bar{p} c_3\big(c_3+c_4\big)+
2(\bar{p}+1)^2c_2c_3-\bar{p}(\bar{p}+1) \big(c_1+c_2\big)c_2\big]	.	
\end{align}
\end{subequations}
\end{widetext}
The success probability for ED round $k$ is given by
\begin{align}
P_\text{ED}^{\prime\,(k)}=(c_1+c_4)^2+(c_2+c_3)^2 + \bar{p} (2c_1+2c_4-1)^2.
\end{align}
\noindent\paragraph{Entanglement swapping.} Similar to Eqs.~\eqref{EDtrafo_hqr}, we neglect the index for the previous nesting level $n-1$. 
The Bell coefficients transform under the ES protocol according to
\begin{widetext}
\begin{subequations}
\label{EStrafo_hqr}
\begin{align}
c_{1,n}&= 2\big(c_1c_4+c_2c_3\big)+2p_c\big(c_1(1-c_1-3c_4)-c_2(c_3-c_4)-(c_2-c_4)c_3\big)+p_c^2\big(2c_1+2c_4-1\big)^2\\ 
c_{2,n}&= 2\big(c_1c_3+c_2c_4\big)+p_c\big((2c_1+2c_4-1)^2 + 2(c_1-c_4)(c_2-c_3)\big) -p_c^2\big(2c_1+2c_4-1\big)^2 , \\
c_{3,n}&= 2\big(c_1c_2+c_3c_4\big)+p_c\big((2c_1+2c_4-1)^2 - 2(c_1-c_4)(c_2-c_3)\big) -p_c^2\big(2c_1+2c_4-1\big)^2 , \\ 
c_{4,n}&= \sum\limits_{i=1}^4c_i^2 -2p_c\Big(\sum\limits_{i=1}^4c_i^2 - (c_1+c_4)(c_2+c_3) \Big) + p_c^2\big(2c_1+2c_4-1\big)^2 . 
\end{align}
\end{subequations}
\end{widetext}
\bibliography{TimoHolz_DIQKD_analysis}
\end{document}